\documentclass{article}

\usepackage{amsthm,amsmath,amssymb}
\usepackage{xspace}
\usepackage{epsfig,wrapfig}
\usepackage{tabularx}                  

\usepackage[latin1]{inputenc}

\usepackage{setspace,color}
\usepackage{multirow}

\newtheorem{theorem}{Theorem}
\newtheorem{lem}[theorem]{Lemma}
\newtheorem{cor}[theorem]{Corollary}
\newtheorem{definition}{Definition}

\usepackage[labelfont=bf,font=small,indention=0.3cm,margin=0cm]{caption}

\usepackage[numbers,sort&compress,longnamesfirst,sectionbib]{natbib}

\usepackage[ruled]{algorithm2e}
\usepackage{algorithmic}

\newcommand{\Rd}{\mathbb{R}^d}
\newcommand{\Rdp}{\mathbb{R}_{\ge 0}^d}
\newcommand{\R}{\mathbb{R}}

\newcommand{\Oh}[1]{\mathcal{O}(#1)}
\newcommand{\Ohbig}[1]{\mathcal{O}\big(#1\big)}
\newcommand{\OhBig}[1]{\mathcal{O}\Big(#1\Big)}

\newcommand{\Ohstar}[1]{\mathcal{O}^*(#1)}
\newcommand{\Ohstarbig}[1]{\mathcal{O}^*\big(#1\big)}
\newcommand{\fun}[1]{\textsc{#1}}
\newcommand{\Set}{\mathcal{S}}
\newcommand{\Vol}[1]{\textsc{vol}(#1)}
\newcommand{\Volbig}[1]{\textsc{vol}\big(#1\big)}

\newcommand{\cov}{\textsc{cov}}
\newcommand{\uc}{\ensuremath{\tilde{C}}}
\newcommand{\la}{47}
\newcommand{\Ex}{\textup{Ex}}

\newcommand{\tabref}[1]{Table~\ref{tab:#1}}
\newcommand{\defref}[1]{Definition~\ref{def:#1}}
\newcommand{\thmref}[1]{Theorem~\ref{thm:#1}}
\newcommand{\lemref}[1]{Lemma~\ref{lem:#1}}

\newcommand{\corref}[1]{Corollary~\ref{cor:#1}}
\newcommand{\secref}[1]{Section~\ref{sec:#1}}

\newcommand{\eq}[1]{equation~\eqref{eq:#1}}

\newcommand{\ee}{\varepsilon}
\newcommand{\eeP}{\ensuremath{\varepsilon_{\textup{P}}}}
\newcommand{\eeV}{\ensuremath{\varepsilon_{\textup{V}}}}
\newcommand{\eeS}{\ensuremath{\varepsilon_{\textup{S}}}}
\newcommand{\eetil}{\tilde{\varepsilon}}

\renewcommand{\P}{\textup{\textbf{P}}\xspace}
\newcommand{\NP}{\textup{\textbf{NP}}\xspace}
\newcommand{\NPhard}{\textup{\textbf{NP}}\nobreakdash-hard\xspace}
\newcommand{\BPP}{\textup{\textbf{BPP}}\xspace}
\newcommand{\Wone}{\textup{\textbf{W[1]}}\xspace}
\newcommand{\FPT}{\textup{\textbf{FPT}}\xspace}

\newcommand{\SP}{\textup{\textbf{\#P}}}

\newcommand{\SPhard}{\SP\nobreakdash-hard\xspace}
\newcommand{\SPhardness}{\SP\nobreakdash-hardness\xspace}
\newcommand{\SMONCNF}{\textup{\#MON\nobreakdash-CNF}\xspace}
\newcommand{\APX}{\textup{APX}\xspace}
\newcommand{\APXhard}{\textup{APX}\nobreakdash-hard\xspace}
\newcommand{\SDNF}{\textup{\#DNF}\xspace}
\newcommand{\SSAT}{\textup{\#SAT}\xspace}

\newcommand{\oB}{\,\overline{\!B}}

\newcommand{\KMP}{\textup{KMP}\xspace}

\newcommand{\ApproxUnion}{\fun{ApproxUnion}\xspace}
\newcommand{\PointQuery}{\fun{PointQuery}\xspace}
\newcommand{\VolumeQuery}{\fun{VolumeQuery}\xspace}
\newcommand{\SampleQuery}{\fun{SampleQuery}\xspace}

\newcommand{\epsapproximation}{$\ee$\nobreakdash-approximation\xspace}

\let\oldsqrt\sqrt
\def\hksqrt{\mathpalette\DHLhksqrt}
\def\DHLhksqrt#1#2{\setbox0=\hbox{$#1\oldsqrt{#2\,}$}\dimen0=\ht0
   \advance\dimen0-0.2\ht0
   \setbox2=\hbox{\vrule height\ht0 depth -\dimen0}%
   {\box0\lower0.4pt\box2}}
\renewcommand\sqrt\hksqrt
\renewcommand{\leq}{\leqslant}
\renewcommand{\geq}{\geqslant}
\renewcommand\epsilon\varepsilon

\catcode`@=11
\def\nphantom{\v@true\h@true\nph@nt}
\def\nvphantom{\v@true\h@false\nph@nt}
\def\nhphantom{\v@false\h@true\nph@nt}
\def\nph@nt{\ifmmode\def\next{\mathpalette\nmathph@nt}%
  \else\let\next\nmakeph@nt\fi\next}
\def\nmakeph@nt#1{\setbox\z@\hbox{#1}\nfinph@nt}
\def\nmathph@nt#1#2{\setbox\z@\hbox{$\m@th#1{#2}$}\nfinph@nt}
\def\nfinph@nt{\setbox\tw@\null
  \ifv@ \ht\tw@\ht\z@ \dp\tw@\dp\z@\fi
  \ifh@ \wd\tw@-\wd\z@\fi \box\tw@}

\begin{document}

\title{Approximating the volume of
       unions and intersections
       of high-dimensional geometric objects}

\author{Karl Bringmann \and Tobias Friedrich}


\maketitle

\begin{abstract}
    We consider the computation of the volume of the union of high-dimensional
    geometric objects.  While showing that this problem is \SPhard
    already for very simple bodies,
    we give a fast FPRAS for all objects where one can (1) test
    whether a given point lies inside the object, (2) sample a point
    uniformly, and (3) calculate the volume of the object in polynomial time.
    It suffices to be able to answer all three questions approximately.
    We show that this holds for a large class of objects.
    It implies that Klee's measure problem can be approximated
    efficiently even though it is \SPhard and hence
    cannot be solved exactly in time polynomial in the number of dimensions unless $\P=\NP$.
    Our algorithm also allows to efficiently approximate
    the volume of the union of convex bodies given by weak membership oracles.

    For the analogous problem of the intersection of high-dimensional
    geometric objects we prove \SPhardness for boxes and
    show that there is no multiplicative polynomial-time $2^{d^{1-\ee}}$\!\nobreakdash-approximation
    for certain boxes unless $\NP=\BPP$, but give a simple additive polynomial-time \epsapproximation.
\end{abstract}

\sloppy


\section{Introduction}
\label{sec:intro}

Given $n$ bodies in the $d$\nobreakdash-dimensional space, how efficiently can we compute
the volume of the union and the intersection?  We consider this basic geometric problem
for different kinds of bodies.
The tractability of this problem
highly depends on the representation and the complexity of the given objects. For
many classes of objects already computing the volume of one body can be hard.
For example, calculating the volume of a polytope given either as a list of vertices or
as a list of facets is \SPhard~\cite{DyerF88,Khachiyan89}.
For convex bodies given by a membership oracle one can also show that even though
there can be no deterministic $(\Oh1 d / \log d)^d$\nobreakdash-approximation for $d \ge 2$~\citep{BaranyF87},
one can still approximate the volume by an FPRAS (fully polynomial-time randomized approximation scheme).
In a seminal paper \citet{DyerFK91} gave an $\Ohstar{d^{23}}$
algorithm, 
which was subsequently improved in a series of papers~\citep{LovaszS90,ApplegateK91,LovaszS93,KannanLS97}
to $\Ohstar{d^4}$~\citep{LovaszV06} (where the asterisk hides powers of the approximation ratio and $\log d$).

Volume computation of unions can be hard not only for bodies whose
volume is hard to calculate.  One famous example for this is \emph{Klee's Measure Problem} (KMP).
Given $n$ axis-parallel boxes in the $d$\nobreakdash-dimensional space ($d$ constant),
the problem asks for the measure of their union.
In 1977, Victor Klee showed that it can be solved in time $\Oh{n \log n}$ for $d=1$~\citep{Klee77}.
This was generalized to $d>1$ dimensions by \citet{Bentley77} in the same year.
He presented an algorithm which runs in $\Oh{n^{d-1} \log n}$,
which was later improved by \citet{LeeuwenW81} to $\Oh{n^{d-1}}$.
In 1988, \citet{OvermarsY91} obtained an $\Oh{n^{d/2} \log n}$ algorithm.
This was the fastest algorithm for $d\geq3$ until very recently \citet{Chan09}
presented a slightly improved version of \citeauthor{OvermarsY91}'s algorithm that
runs in time $n^{d/2} \, 2^{\Oh{\log^*n}}$,
where $\log^*$ denotes the iterated logarithm.
So far, the only known lower bound is $\Omega(n \log n)$ for any~$d$~\citep{FredmanW78}.
\citet{Chan09} also proves that no algorithm of runtime $n^{o(d)}$ is possible
assuming $\Wone\neq\FPT$, which is a weaker result than $\P\neq\NP$ and a
commonly accepted conjecture on fixed-parameter tractability.
Note that the worst-case combinatorial complexity (i.e., the number of faces of all dimensions on the boundary
of the union) of $\Theta(n^d)$ does not imply any bounds on the computational complexity.
There are various algorithms for special cases, e.g., for
hypercubes~\citep{AgarwalKS07,KaplanRSV07} and unit hypercubes~\citep{Chan03}.
In this paper we explore the opposite direction and
examine the union of more general geometric objects.

\subsection*{Our results}

It is not hard to see that \KMP is \SPhard (see \thmref{hardness}).
Hence it cannot be solved in time polynomial in the number of dimensions unless $\P = \NP$.
This shows that \emph{exact} volume
computation of unions is intractable for all classes of bodies that contain
axis-parallel boxes.
This motivates the development of \emph{approximation algorithms} for the volume
computation of unions.
Based on an FPRAS for \#DNF by \citet{KarpLM89},
we give an FPRAS for a large class of bodies including boxes, spheres,
polytopes, convex bodies determined by an oracle, and schlicht domains.
Additionally, also fixed affine transformations of the forementioned objects can be allowed.
The underlying bodies~$B$ just have to support the following oracle queries in polynomial time:
\begin{itemize}
    \setlength{\itemsep}{0pt}
    \setlength{\parskip}{0pt}
    \item $\PointQuery(x,B)$: Is point $x\in\Rd$ an element of body $B$?
    \item $\VolumeQuery(B)$: What is the volume of body $B$?
    \item $\SampleQuery(B)$: Return a random uniformly distributed point $x\in B$.
\end{itemize}
\PointQuery is a very natural condition which is fulfilled in almost all practical cases.
The \VolumeQuery condition is important as it could
be the case that no efficient approximation of the volume of one of the bodies
itself is possible.  This, of course, prevents an efficient approximation of the union of such bodies.
The \SampleQuery is crucial for our FPRAS.
In \secref{otherobj} we will show that it is efficiently computable for a wide range of bodies.

An important feature of our algorithm is that it suffices that all three oracles are weak.
More precisely, we allow the following
\emph{relaxation} for every body~$B$ ($\Vol{B}$ denotes the volume of a body~$B$
in the standard Lebesgue measure on~$\R^d$,
more details are given in \secref{union}):
\begin{itemize}
    \setlength{\itemsep}{0pt}
    \setlength{\parskip}{0pt}
    \item $\PointQuery(x,B)$ answers true if and only if $x\in B'$ for a fixed $B'\subset \Rd$
          with $\Vol{ (B'\setminus B) \cup (B\setminus B') } \leq \eeP \Vol{B}$.
    \item $\VolumeQuery(B)$ returns a value $V'$ with $(1-\eeV)\,\Vol{B}\leq V' \leq (1+\eeV)\,\Vol{B}$.
    \item $\SampleQuery(B)$ returns only an \emph{almost} uniformly distributed random point~\citep{KannanLS97},
          that is, it suffices to get a random point $x \in B'$ (with $B'$ as above)
          such that for the
          probability density $f$ we have for every point $x$: $|f(x) - 1/\Vol{B'}| < \eeS$.
\end{itemize}
Let $P(d)$ be the worst \PointQuery runtime\footnote{The runtime of the oracles can also depend
on the required approximation guarantees. In order to simplify the notation,
this dependency is not made explicit.}
of any of our bodies,
analogously $V(d)$ for \VolumeQuery,
and $S(d)$ for \SampleQuery.
Then our FPRAS has a runtime of
$\Ohbig{n V(d) + \frac{n}{\ee^2} \,(S(d) + P(d))}$ for producing an
\epsapproximation\footnote{We will
always assume that $\ee$ is small, that is, $0<\ee<1$.}
with probability $\ge \tfrac{3}{4}$
if the errors
 of the underlying oracles are small, i.e., $\eeS, \eeV \le \frac{\ee^2}{47 n}$ and $\eeP \le \frac{\ee^2}{47 n^2}$.
For example for boxes (e.g., for \KMP), this reduces to $\Ohbig{\frac{d n}{\ee^2}}$
and is the first FPRAS for this problems.
In \secref{otherobj} we also show that our algorithm is an FPRAS for the
computation of the volume of the union of convex bodies.

The canonical next question is the computation of the volume of the
\emph{intersection of bodies} in~$\Rd$. It is clear that most of the problems from
above apply to this question, too. \SPhardness for general, i.e.,
not necessarily axis-parallel, boxes follows directly from the hardness of
computing the volume of a polytope~\cite{DyerF88,Khachiyan89}.
This leaves open whether there are efficient approximation algorithms
for the volume of intersection.
In \secref{intersection} we show that there cannot be a (deterministic or randomized) multiplicative
$2^{d^{1-\ee}}$\nobreakdash-approximation in general, unless $\NP = \BPP$
by identifying a hard subproblem.
Instead we give an additive \epsapproximation, which is therefore the best we can hope for.
It has a runtime of $\Ohbig{n \, V(d) + \ee^{-2} \, S(d) + n \, \ee^{-2}\, P(d)}$,
which gives $\Ohbig{\frac{d \, n}{\ee^2}}$ for boxes.


\section{Volume computation of unions}
\label{sec:union}

In this section we show that the volume computation of unions is \SPhard already for
very simple axis-parallel boxes.
After that we give an FPRAS for approximating the volume of the union of bodies which satisfy the
three aforementioned oracles and describe several classes of objects for which
the oracles can be answered efficiently.

\subsection{Computational complexity of union calculations}
\label{sec:hardness}

Consider the following problem: Let $\Set$ be a set of $n$ axis-parallel boxes in
$\Rd$ of the form $B = [a_1,b_1]\times \cdots \times [a_d,b_d]$ with $a_i, b_i
\in \R, a_i < b_i$. The volume of one such box is $\Vol{B} =
\prod_{i=1}^d (b_i-a_i)$. To compute the volume of the union of these boxes is
known as Klee's Measure Problem (\KMP).

It is know that the associated decision problem of deciding whether
there is a point that is not in the union is \NPhard~\citep{Chan09}.
We consider the actual counting problem and prove in the
following \thmref{hardness} that \KMP is \SPhard.
To the best of our knowledge there is no published result that explicitly states that \KMP is \SPhard.
However, without mentioning this implication,
\citet{SuzukiI04} sketch a reduction from \#SAT to \KMP.
We present a reduction from \SMONCNF to \KMP which
we can reuse in \thmref{noapprox} for the hardness proof for intersections.
\SMONCNF counts the number of satisfying assignments of a
Boolean formula in conjunctive normal form in which all variables are unnegated.
While the problem of deciding satisfiability of such formula is trivial,
counting the number of satisfying assignments is \SPhard~\citep{Roth96}.

\begin{theorem}
    \label{thm:hardness}
    \KMP is \SPhard.
\end{theorem}
\begin{proof}
    To reduce \SMONCNF to \KMP, let
    $
        f = \bigwedge_{k=1}^n \bigvee_{i \in C_k} x_i
    $
    be a monotone Boolean formula given in CNF
    with $C_k \subset [d]:=\{1,\ldots,d\}$, for $k\in[n]$,
    $d$ the number of variables,
    $n$ the number of clauses. Since the
    number of satisfying assignments of $f$ is equal to $2^d$ minus the number
    of satisfying assignments of its negation, we instead count the latter:
    Consider the negated formula $\bar{f} = \bigvee_{k=1}^n \bigwedge_{i \in
    C_k} \neg x_i$. First, we construct a box $A_k = [0,q_1^{(k)}]\times \cdots
    \times[0,q_d^{(k)}]$ in~$\Rd$ for each clause $C_k$ with one vertex at the
    origin and the opposite vertex at $(q_1^{(k)},\ldots,q_d^{(k)})$, where we set
    \[
        q_i^{(k)} =
        \left\{
        \begin{aligned}
            & 1, \quad \text{if} \quad i \in C_k \\
            & 2, \quad \text{if} \quad i \not\in C_k
        \end{aligned}
        \right.
        , \quad i\in[d].
    \]

    \noindent
    Observe that the union of the boxes $A_k$ can be written as a union of boxes
    of the form $U_x = [x_1, x_1+1]\times \cdots \times [x_d,
    x_d+1]$ with $x=(x_1,\ldots,x_d)\in\{0,1\}^d$.
    Let $x\in\{0,1\}^d$ and $U_x\subseteq\bigcup_{k=1}^n A_k$.
    Then there is a $k$ such that $q_i^{(k)} = 2$ for all~$i$ with $x_i=1$.
    By definition of $q_i^{(k)}$, this implies that
    $\bigwedge_{i \in C_k} \neg x_i$ and also $\bar{f}$ are satisfied.

    The same holds in the opposite direction, that is, if
    $x$ satisfies $\bar{f}$ then
    $U_x\subseteq\bigcup_{k=1}^n A_k$.
    Hence, since
    $\Vol{U_x} = 1$, we have $\Vol{\bigcup_{k=1}^n A_k} =
    |\{x \in \{0,1\}^d \mid x \: \text{satisfies}
    \: \bar{f}\}|$. Thus a polynomial time algorithm for \KMP would result in a
    polynomial time algorithm for \SMONCNF, which proves the claim.
\end{proof}

Note that we actually proved a little bit more than stated in the theorem.
That is, we proved that even calculating the volume of the union of boxes
which all have the point $0^d$ in common is \SPhard.
This specific problem is known as \emph{hypervolume indicator}~\citep{ZitzlerT99}
and is a very popular and widely used measure of fitness of Pareto sets
in evolutionary multi-objective optimization.


\subsection{Approximation algorithm for the volume of unions}
\label{sec:approx}

In this section we present an FPRAS for computing the volume of the union of objects for which we can
answer \PointQuery, \VolumeQuery, and \SampleQuery in polynomial time.
The input of our algorithm \ApproxUnion are the approximation ratio $\ee$
and the bodies $B_1,\ldots,B_n$ in~$\R^d$ defined by the three oracles.
It computes an approximation $\tilde{U} \in \R$ of
$U:=\Volbig{\bigcup_{i=1}^n B_i}$
such that
\begin{equation}
    \label{eq:FPRASapprox}
    \Pr\big[ (1-\ee)\,U
        \le \tilde{U}
        \le (1+\ee)\,U \big]
    \ge \tfrac{3}{4}
\end{equation}
in time polynomial in~$n$, $1/\ee$ and the query runtimes.
Note that the constant $\tfrac{3}{4}$ can be increased to any number
by using a probability amplification technique.

\begin{algorithm}[t]
\label{alg}
\caption{\ApproxUnion$(\Set, \ee, \eeP, \eeV, \eeS)$ calculates an
         $\ee$-approximation of $U = \Vol{\bigcup_{i=1}^n B_i}$ for a set of bodies $S = \{B_1,\ldots,B_n\}$
         in~$\Rd$ determined by the oracles $\PointQuery$, $\VolumeQuery$
         and $\SampleQuery$ with error ratios $\eeP,\eeV,\eeS$.}
\begin{algorithmic}
    \STATE $\eetil := \frac{\ee-\eeV}{1+\eeV}$
    \STATE $\uc := \frac{(1+\eeS)\,(1+\eeV)\,(1+n \eeP)}{(1-\eeV)\,(1-\eeP)}$
    \STATE $T := \frac{24 \ln(2) \,(1+\eetil) n}{\eetil^2 - 8 \,(\uc - 1)\, n}$
    \FORALL{$i \in [n]$}
        \STATE compute $V_i' := \VolumeQuery(B_i)$
    \ENDFOR
    \STATE $V' := \sum_{i=1}^n V_i'$
    \FOR{$M:=0$ \textbf{to} $\infty$}
         \STATE choose $i \in [n]$ with probability $V_i' / V'$
         \STATE $x := \SampleQuery(B_i)$
         \STATE $t_M := 0$
         \REPEAT
              \STATE \textbf{if} $t_0+\ldots+t_M \ge T$ \textbf{then return} $\frac{T \, V'}{n \, M}$
              \STATE choose $j \in [n]$ uniformly at random
              \STATE $t_M := t_M + 1$
         \UNTIL{\PointQuery$(x,B_j)$}
    \ENDFOR
\end{algorithmic}
\end{algorithm}

We are following the algorithm of \citet{KarpLM89} which the authors used for
approximating \#DNF and other counting problems on discrete sets. The two main
differences are that here we are handling continuous bodies in~$\Rd$ and
that we allow erroneous oracles.
The latter relaxation is crucial to incorporate, amongst other things, the class of convex bodies.
Our algorithm \ApproxUnion is shown on page~\pageref{alg}.
The total number of steps of the algorithm is $T$.
This number is chosen in advance such that one can prove that
\eq{FPRASapprox} holds.
The algorithm itself is very simple.
First, it queries the volumes\footnote{Note that here and in the remainder an
unprimed variable denotes an exact value and
a primed variable denotes a value subject to some error introduced by the erroneous oracles.}
$V_i'$ of the bodies $B_i$ and computes $V' = \sum_{i=1}^n V_i'$.
Then it repeats the following:
It chooses a body $B_i$ with probability (roughly\footnote{``roughly'' only
for erroneous oracles.}) proportional to its volume
and chooses a point $x\in B_i$ (roughly\footnotemark[4]) uniformly at random.
Afterwards, the algorithm chooses bodies $B_j$ with probability $1/n$
and (roughly\footnotemark[4]) checks whether $x\in B_j$.
The number $t_M$ of chosen bodies until we find a $B_j$ with $x \in B_j$
can then be used to estimate how many bodies $B_j$ cover $x$.

For this, observe that the point $x$ is chosen
with probability density $k(x)/V'$ with
$k(x)= | i\in[n] \mid x\in B_i |$.
Each number $t_M$ has expected value (roughly\footnotemark[4])
$n/k(x)$ for a fixed $x \in \bigcup_i B_i$.
Hence when \PointQuery$(x,B_j)$ answers yes,
$t_M$ should be of the order of (roughly\footnotemark[4])
$
\int_x n/k(x) \cdot k(x)/V' \,dx
   = \int_x (n/V') \,dx
  = nU/V'$.
This implies that when $\tilde{U}$ is returned and
$T=t_0+\ldots+t_M$, the value
$n\,U\,M/V'$ is near $T$, i.e., $T\,V'/n\,M$ is near $U$.
This gives the intuition why
the algorithm returns the approximation $\tilde{U}=\frac{T \, V'}{n \, M}$.

In \secref{analysis} we show correctness of \ApproxUnion.
More precisely, we show that it returns an \epsapproximation
with probability $\ge \tfrac{3}{4}$
and $T = \Ohbig{\frac{n}{\ee^2}}$
if $\eeS, \eeV \le \frac{\ee^2}{47 n}$ and $\eeP \le \frac{\ee^2}{47 n^2}$.
The last inequality reflects the fact that we cannot be arbitrarily
accurate if the given oracles are inaccurate.
If all oracles can be calculated accurately, i.e., if $\eeP = \eeS = \eeV = 0$,
the algorithm runs for just
$T = \frac{8 \ln(8) \,(1 + \ee) \,n}{\ee^2}$
many steps.

Overall, the runtime of \ApproxUnion is clearly
$
    \Oh{n \cdot V(d) + M \cdot S(d) + T \cdot P(d)}
    = \Oh{n \cdot V(d) + T \cdot (S(d) + P(d) )},
$
where $V(d)$ is the worst \VolumeQuery time for any of the bodies,
analogously $S(d)$ for \SampleQuery and $P(d)$ for \PointQuery.
If $\eeS, \eeV \le \frac{\ee^2}{47 n}$ and $\eeP \le \frac{\ee^2}{47 n^2}$,
the runtime is $\Ohbig{n \, V(d) + \frac{n}{\ee^2} \, (S(d) + P(d) )}$.

For boxes all three oracles can be computed exactly in~$\Oh{d}$.
This implies that our algorithm \ApproxUnion gives an \epsapproximation
of \KMP with probability $\ge \tfrac{3}{4}$
in runtime $\Ohbig{\frac{n d}{\ee^2}}$.
For more complex objects like convex bodies determined by a membership oracle,
there are no exact oracles.  The following section discusses different classes
of objects for which our algorithm can be applied.


\subsection{Classes of objects supported by our FPRAS}
\label{sec:otherobj}

Finding an FPRAS for the union of a certain class of geometric objects
now reduces to calculating the respective \PointQuery, \VolumeQuery and \SampleQuery in polynomial time.
We assume that we can get a random real number in constant time.
Then all three oracles can be calculated in time $\Oh{d}$ for $d$\nobreakdash-dimensional boxes.
This already yields an FPRAS
for the volume of the union of arbitrary boxes, e.g., for \KMP.
Note that if we have a body for which we can answer all those queries, all
affine transformations of this body fulfill these three oracles, too.
We will now present three further classes of geometric objects.

\paragraph{Generalized spheres and boxes} Let
$\textbf{B}_k$ be the class of boxes of dimension~$k$, i.e., $\textbf{B}_k =
\{[a_1,b_1]\times\cdots\times[a_k,b_k] \mid a_i,b_i \in \R, a_i < b_i \}$ and
$\textbf{S}_\ell$ the class of spheres of dimension $\ell$.
We can combine any box $B \in \textbf{B}_k$ and
sphere $S \in \textbf{S}_{d-k}$ to get a
$d$\nobreakdash-dimensional object $B \times S$.
Furthermore, we can permute the dimensions
afterwards to get a generalized ``box-sphere.''
in~$\R^3$ this corresponds to boxes, spheres and cylinders.
\VolumeQuery can be computed easily as
we can compute the volume of a sphere by a well-known formula
and thus the volume of the product $B \times S$.
As one can check whether a given point $x = (x_1,\ldots,x_d)$ lies in $B \times S$ by checking
whether $(x_1,\ldots,x_k)$ lies in~$B$ and $(x_{k+1},\ldots,x_d)$ lies in~$S$,
also \PointQuery is a standard task of geometry.
To answer \SampleQuery, it suffices to choose a random point $(x_1,\ldots,x_k)$ in~$B$
and to choose a random point inside the sphere $S$, which can be done in polynomial time
as described, e.g., by \citet{Muller59}.

\paragraph{Convex bodies}  As mentioned in the introduction, exact calculation
of \VolumeQuery for a polytope given as a list of vertices or
facets is \SPhard~\cite{DyerF88,Khachiyan89}.
Since there are randomized
approximation algorithms (see \citet{DyerFK91} for the first one)
for the volume of a convex body
determined by a membership oracle,
we can answer \VolumeQuery approximately.
The same holds for \SampleQuery as these algorithms make
use of an almost uniform sampling method on convex bodies.
See \citet{LovaszV06} for a result showing that \VolumeQuery can be answered with
$\Ohstarbig{\frac{d^4}{\eeV^2}}$ questions to the membership oracle and
\SampleQuery with $\Ohstarbig{\frac{d^3}{\eeS^2}}$ queries,
for arbitrary errors $\eeV, \eeS > 0$ (where the asterisk
hides factors of $\log(d)$ and $\log(1/\eeV)$ or $\log(1/\eeS)$).
\PointQuery can naturally be answered with a single question to the membership oracle.
By choosing $\eeV = \eeS = \frac{\ee^2}{47 n}$,
\thmref{FPRASproof} together with \lemref{cond} shows that \ApproxUnion is an FPRAS for the
volume of the union of convex bodies which uses
$\Ohstarbig{\frac{n^3 d^3}{\ee^4}(d+\frac{1}{\ee^2})}$ membership queries.

\paragraph{Star-shaped bodies}
Star-shaped bodies are a generalization of convex bodies which
have at least one point such that every line through the point
has a convex intersection with the body.
They can also be viewed as the union of convex sets,
with all the convex sets having a nonempty intersection.
The subset of points that can ``see''
the full set is called the kernel of the star-shaped set.
Assuming that we are given membership oracles for the body
as well as for the kernel,
\citet{CDV10} recently showed that for star-shaped bodies
\SampleQuery can be answered with $\Ohstarbig{\frac{d^3}{\eta^3\,\eeS^2}}$ 
questions to the membership oracle, where $\eta$
is the fraction of the volume taken up by the kernel.
We can also approximate the volume of the (convex) kernel 
with $\Ohstarbig{\frac{d^4}{\eeV^2}}$ questions to the membership oracle
as discussed above
and estimate $\eta$ by $\Oh{\frac{1}{\eta^2\,\eeV^2}}$ samples.
Then the runtime of \VolumeQuery is 
$\Ohstarbig{\frac{d^4}{\eeV^2} + \frac{d^3}{\eta^5\,\eeV^2\,\eeS^2}}$.
\PointQuery can again naturally be answered with a single question to the membership oracle.

\paragraph{Schlicht domains} Let $a_i, b_i\colon \R^{i-1} \rightarrow \R$ be
continuous functions with $a_i \le b_i$, where $a_1, b_1$ are constants.
Let $K \subset \Rd$ be defined as the set of all points $(x_1,\ldots,x_d) \in
\Rd$ such that
$a_1 \le x_1 \le b_1, a_2(x_1) \le x_2 \le b_2(x_1), \ldots,
a_d(x_1,\ldots,x_{d-1}) \le x_d \le b_d(x_1,\ldots,x_{d-1})$.
$K$ is called a schlicht domain in functional analysis.
Fubini's theorem for schlicht domains states that we can integrate a function $f\colon K \rightarrow \R$
by iteratively integrating first over $x_d$, then over $x_{d-1}$, \ldots, until
we reach $x_1$. This way, by integrating $f(\cdot)=1$, we can compute the volume of a schlicht domain
as long as the integrals are computable in polynomial time, and thus answer \VolumeQuery.
Similarly, we can choose a random uniformly distributed point
inside~$K$: Let $K(y) = \{ (x_1,\ldots,x_d) \in K \mid x_1 = y \}$. Then $K(y)$
is another schlicht domain for every $a_1 \le y \le b_1$. Assume that we can
determine the volume of every such $K(y)$ and the integral $I(y) =
\int_{a_1}^y K(x) \,dx$. Then the inverse function $I^{-1}\colon [0,V] \rightarrow \R$,
where $V = \int_{a_1}^{b_1} K(x) \,dx$ is the volume of~$K$,
allows us to choose a $y$ in $[a_1,b_1]$ with probability proportional to $\Vol{K(y)}$.
By this we can iteratively choose a value $y$ for
$x_1$ and recurse to find a uniformly random point $(y_2,\ldots,y_d)$ in~$K(y)$,
plugging both together to get a uniformly distributed point $(y_1,\ldots,y_d)$ in~$K$.
Hence, as long as we can compute the involved integrals and inverse functions
(or at least approximate them good enough), we can answer \SampleQuery.
 Since \PointQuery is trivially computable --
as long as we can evaluate $a_i$ and $b_i$ efficiently --
this gives an example showing
that the classes of objects that fulfill our three conditions include not only
convex bodies, but also certain schlicht domains.

\medskip
Note that all above mentioned classes of geometric objects contain boxes and
hence our hardness results still hold and an \epsapproximation algorithm
is the best one can hope for (unless $\P=\NP$).


\subsection{Analysis of our algorithm}
\label{sec:analysis}

We now show correctness of our algorithm \ApproxUnion
described in \secref{approx} and prove bounds for its approximation ratio.
The following theorem is our main result for \ApproxUnion.
It holds for exact and weak oracles.
\begin{theorem}
    \label{thm:FPRASproof}
    Given errors $0 \le \eeP, \eeS, \eeV < 1$ of the queries,
    the algorithm $\ApproxUnion(\{B_1,\dots,B_n\}, \ee, \eeP, \eeV, \eeS)$ returns a value $\tilde{U}$ with
    \[
        \Pr\big[ (1-\ee)\,U \le \tilde{U} \le (1+\ee)\,U \big]
        \ge \tfrac{3}{4}
    \]
    choosing
    \[
        T = \frac{24 \ln(2) \,(1+\eetil) \,n}{\eetil^2 - 8 \,(\uc - 1) \,n},
    \]
    under the condition
	\begin{align}
		\label{eq:eeCond}
		\ee > \eeV + 2(1 + \eeV)\sqrt{2(\uc-1)\,n}
	\end{align}
    where
    $U:=\Volbig{\bigcup_{i=1}^n B_i}$,
    $\eetil := \frac{\ee-\eeV}{1+\eeV}$,
    and $\uc := \frac{(1+\eeS)\,(1+\eeV)\,(1 + n \eeP)}{(1-\eeV)\,(1-\eeP)}$.
\end{theorem}

First note that with accurate oracles, i.e., if $\eeP=\eeS=\eeV=0$,
we get $\eetil = \ee$, $\uc = 1$ and, thus, $T = \frac{24 \ln(2) \,(1+\ee) \,n}{\ee^2}$.
As the condition~\eqref{eq:eeCond} becomes trivial, above theorem implies
that our algorithm is indeed an FPRAS.

Given non-zero query errors, one clearly cannot be arbitrary accurate, which is
reflected by the lower bound~\eqref{eq:eeCond} for $\ee$.  However, the following lemma shows
that condition~\eqref{eq:eeCond} is fulfilled for small enough $\eeP$, $\eeS$ and $\eeV$.
\begin{lem}
    \label{lem:cond}
	For $\eeS, \eeV \le \ee^2/(\la n)$ and $\eeP \le \ee^2/(\la n^2)$ the
	condition~\eqref{eq:eeCond} of \thmref{FPRASproof}
	is fulfilled and we have $T = \Ohbig{\frac{n}{\ee^2}}$.
\end{lem}
\begin{proof}
	As the right hand side of \eqref{eq:eeCond} is increasing in~$\eeP$, $\eeV$ and $\eeS$,
	we can assume w.l.o.g.\ that $\eeV = \eeS = \frac{\ee^2}{\la n}$ and $\eeP = \frac{\ee^2}{\la n^2}$.
	This gives $\uc \le \big( 1 + \frac{\ee^2}{\la n} \big)^3 \big( 1 - \frac{\ee^2}{\la n} \big)^2$.
	Observe that $\frac{\ee^2}{\la n} \le \frac{1}{\la}$. Since
	$(1+x)^3 \, (1-x)^{-2} \le 1 + \alpha x$ holds for $\alpha := \frac{8081}{1521}$
	and also $0 \le x \le \frac{1}{40}$, we have $\uc \le 1 + \alpha \frac{\ee^2}{\la n}$.
	Hence, we can upper bound the right hand side of \eqref{eq:eeCond} by
	\begin{align*}
		\eeV + 2\,(1 + \eeV)\sqrt{2(\uc-1)n}
		&\le \tfrac{\ee^2}{\la n} +
		     2\,\left(1 + \tfrac{\ee^2}{\la n}\right)\sqrt{2 \alpha \tfrac{\ee^2}{\la}} \\
		&\le \ee \left(\tfrac{1}{\la} + 2\left(1 + \tfrac{1}{\la}\right) \,\sqrt{\tfrac{2 \alpha}{\la}}\right)
	\end{align*}
	as $1/n \le 1$ and $\ee \le 1$.
	Since $\frac{1}{\la} + 2(1 + \frac{1}{\la}) \,\sqrt{\frac{2 \alpha}{\la}} < 1$,
	condition \eqref{eq:eeCond} is fulfilled.

	We now bound the terms $\eetil$ and $\uc$. Since $\eeV \ge 0$, we clearly have $\eetil \le \ee$ implying
	$1 + \eetil \le 2$. Furthermore, since $\eeV \le \frac{\ee^2}{\la n}$ it also holds that
	\[
		\eetil = \frac{\ee - \eeV}{1+\eeV} \ge \frac{\ee - \frac{\ee^2}{47 n}}{1 + \frac{\ee^2}{47 n}} = \frac{\ee(47n-\ee)}{\ee^2 + 47 n}
		\ge \frac{\ee \, 46 n}{48 n} = \frac{23}{24} \ee
	\]
	where we used $\ee < 1$ and $n \ge 1$. Using this and the upper bound for $\uc$
	we get for the denominator of $T$:
	\[
		\eetil^2-8\,(\uc-1)\,n \ge (\tfrac{23}{24} \ee)^2 - \tfrac{8}{47} \alpha \ee^2 = \tfrac{64375}{4575168} \ee^2.
	\]
	Therefore, we get
	\[
		T = \frac{24 \ln(2)\, (1+\eetil) \,n}{\eetil^2 - 8 \,(\uc - 1) \,n}
		    \le \frac{48 \ln(2)\, n}{\frac{64375}{4575168} \ee^2}
		    < 2365 \frac n{\ee^2}
		    = \OhBig{\frac{n}{\ee^2}}.
		    \qedhere
	\]
\end{proof}

In order to prove \thmref{FPRASproof}, we will generalize the corresponding proof of \citet{KarpLM89}
to cover weak oracles.  For most lemmas it suffices to insert the error constants
$\eeP, \eeS, \eeV$, but for the main proof of \thmref{FPRASproof} one has to be little bit more
careful.

First, recall that we are given bodies $B_1,\dots,B_n$ by oracles, where
$\PointQuery(x,B_i)$ returns true for every $x \in B_i'$, such that the result
is wrong for the set $W_i = (B_i \backslash B_i') \cup (B_i' \backslash B_i)$
with $\Vol{W_i} < \eeP \Vol{B_i}$, which implies
\begin{align}
    \label{eq:VolBiIneq}
    (1-\eeP) \,\Vol{B_i} \le \Vol{B_i'} \le (1+\eeP) \,\Vol{B_i}.
\end{align}
Furthermore, the volume
$V_i'$  of body $B_i$ is computed by \VolumeQuery.
$V_i'$ is an \eeV\nobreakdash-approximation of the corresponding exact volume $V_i$, i.e.,
\begin{align}
    \label{eq:ViIneq}
    (1-\eeV) \,V_i \le V_i' \le (1+\eeV) \,V_i.
\end{align}
We set $V' := \sum_{i=1}^n V_i'$ and $V := \sum_{i=1}^n V_i$. Then it clearly holds that
\begin{align}
    \label{eq:VIneq}
    (1-\eeV) \,V \le V' \le (1+\eeV) \,V.
\end{align}
Furthermore, let $U$ be the exact volume of the union of the $B_i$'s and $\mu = U / V$.

As in \citet{KarpLM89}, we define for a point $x \in \Rd$ the number of covering bodies
$\cov(x) = |\{ i \in [n] \mid \PointQuery(x,B_i) = \text{true} \}|$. Additionally, we set
\[
R_k := \{ x \in \Rd \mid \cov(x) = k \}
\]
and $r_k := \Vol{R_k}$. Then we have $\sum_{k=1}^n k \, r_k = \sum_{i=1}^n \Vol{B_i'}$, so that
\begin{align}
    \label{eq:SumkrkIneq}
    (1-\eeP) \,V \le \sum_{k=1}^n k \, r_k \le (1+\eeP) \,V.
\end{align}
Furthermore, $\sum_{k=1}^n r_k = \Vol{\bigcup_{i=1}^n B_i'}$, so that
\begin{align}
    \label{eq:SumrkIneq}
    (1-n \eeP) \,U \le \sum_{k=1}^n r_k \le (1+n \eeP) \,U.
\end{align}

\noindent
To get a sample point in our algorithm, we first choose an $i \in [n]$ with probability $V_i' / V'$
and then choose a random point $x$ in~$B_i'$ via $\SampleQuery(B_i)$.
We consider the probability of this random point $x$ to lie in the region $R_k$.
With error-free oracles this probability would be $\Pr[x \in R_k] = \frac{k r_k}{V}$ as exactly $k$
bodies cover each point of $R_k$ and we have $\sum_{k=1}^n k r_k = V$ in the error-free setting.
With errors, simple calculations using inequalities \eqref{eq:VolBiIneq}, \eqref{eq:ViIneq} and
\eqref{eq:VIneq} yield
\begin{align}
    \label{eq:PrxRkIneq}
     \frac{(1-\eeS)\,(1-\eeV)}{(1+\eeV)\,(1+\eeP)} \, \frac{k \, r_k}{V}
    \le  \Pr[ x \in R_k ]
    \le  \frac{(1+\eeS)\,(1+\eeV)}{(1-\eeV)\,(1-\eeP)} \, \frac{k \, r_k}{V}.
\end{align}

In the algorithm, $t_m$ denotes the number of iterations in the inner loop
during the $m$\nobreakdash-th iteration of the main loop,
i.e., the number of trials to find a box containing the $m$\nobreakdash-th point~$x$.
These $t_m$ are independent identically distributed random variables. Let $t$
be a variable distributed as each $t_m$. Then the following Lemma holds.

\begin{lem}
\label{lem:Exeht}
    Let $0 \le \lambda \le \frac{1}{2}$, and
    $\uc = \frac{(1+\eeS)\,(1+\eeV)\,(1+n \eeP)}{(1-\eeV)\,(1-\eeP)}$.
    Then
    \begin{align*}
        \Ex[e^{\lambda t / n}] &\le \uc e^{(\lambda + 2\lambda^2)\mu} \text{ and}  \\
        \Ex[e^{-\lambda t / n}] &\le \uc e^{-(\lambda - \lambda^2)\mu}.
    \end{align*}
\end{lem}

These bounds closely match Lemmas 7 and 9 of \citet{KarpLM89}. Adapting the proof is straightforward.
The factor $\uc$ arises by the usage of inequalities~(\ref{eq:SumrkIneq}) and~(\ref{eq:PrxRkIneq}).

In the remainder, let $S_\ell := \sum_{i=0}^\ell t_i$ be
the step at which the $\ell$\nobreakdash-th trial is completed and let $N_i$ be the number
of trials completed after step $i$, so that in the end $M = N_T$.
Then $N_i < \ell$ if and only if $S_\ell > i$.

\begin{cor}
	\label{cor:Sellupper}
	Let $\ee \le 2$. Then
	\begin{align*}
	    \Pr[S_\ell > (1+\ee)\,n \mu \ell] &\le \uc^\ell e^{-\mu \ee^2 \ell / 8} \text{ and}  \\
    	\Pr[S_\ell < (1-\ee)\,n \mu \ell] &\le \uc^\ell e^{-\mu \ee^2 \ell / 8}.
	\end{align*}
\end{cor}

This corresponds to Corollaries~8 and~10 in \citet{KarpLM89}. We can reuse their proof word for word;
the only change is our \lemref{Exeht} which brings in the factor $\uc$.

It remains to prove \thmref{FPRASproof}. The corresponding theorem of \citet{KarpLM89}
is called Self-Adjusting Coverage Algorithm Theorem II. However, adapting their proof is not as straightforward
as for the previous lemmas.  It is presented in more detail in the remainder of this section.

\begin{proof}[Proof of \thmref{FPRASproof}]
    Let $\eetil = \frac{\ee-\eeV}{1+\eeV}$ and
    $\uc = \frac{(1+\eeS)\,(1+\eeV)\,(1+n \eeP)}{(1-\eeV)\,(1-\eeP)}$
    and assume
    \begin{equation}
        \label{eq:eeIneq}
        \ee > \eeV + 2(1 + \eeV)\sqrt{2\,(\uc-1)\,n}.
    \end{equation}
    This implies $\eetil > 0$ and $\eetil^2 - 8 \, (\uc - 1) \,n > 0$.
    Let
    \begin{align*}
        k_1 &:= \frac{24 \ln(2) \,(1+\eetil)}{\mu \, (\eetil^2 - 8 \,(\uc - 1) \,n ) \,(1+\eetil) }
        \intertext{and}
        k_2 &:=  \frac{24 \ln(2) \,(1+\eetil)}{\mu \, (\eetil^2 - 8 \, (\uc - 1) \,n ) \,(1-\eetil) }.
    \end{align*}
    Then $T = k_1 n \mu \, (1+\eetil) = k_2 n \mu \, (1-\eetil)$.
    Now, if we have $k_1 \le M \le k_2$, then
    \[
        \frac{T}{k_2} \le \frac{T}{M} \le \frac{T}{k_1}
    \]
    and thus
    \[
        \frac{T \, V'}{n k_2} \le \frac{T \, V'}{n M} = \tilde{U} \le \frac{T \, V'}{n k_1}.
    \]
    By plugging in~$T$, $k_1$ and $k_2$ we get
    \[
        (1-\eetil) \,\mu V' \le \tilde{U} \le (1+\eetil) \,\mu V'.
    \]
    A little calculus shows that $(1 - \eetil)\,(1 - \eeV) \ge 1 - \ee$, only based on the
    definition of $\eetil$, the non-negativity of $\eeV$ and $\ee$, and $\eeV \le \ee$.
    By using this,
    \eq{VIneq},
    the fact $(1 + \eetil) \,(1 + \eeV) = 1 + \ee$,
    and $\mu = U/V$ we get
    \[
        (1 - \ee) \, U \le \tilde{U} \le (1+\ee) \,U,
    \]
    and thus the estimation is an \epsapproximation, if $k_1 \le M \le k_2$. Hence, it suffices to show
    \begin{equation}
        \label{eq:FPRASproof:suffices}
        \Pr[M > k_2] + \Pr[M < k_1] \le \tfrac{1}{4}.
    \end{equation}
    We have
    \[
        \Pr[M < k_1] = \Pr[S_{k_1} > T]
                     = \Pr[S_{k_1} > k_1 n \mu \, (1+\eetil)].
    \]
    By $\eetil > 0$ we have, using \corref{Sellupper},
    \begin{align*}
    \Pr[M < k_1] &\le \uc^{k_1} e^{-\mu \eetil^2 k_1 / 8}
                 \le e^{(\uc-1) \, k_1} e^{-\mu \eetil^2 k_1 / 8}\\
                 &= e^{(\uc - 1 -\mu \eetil^2 / 8) \, k_1}
                 = e^{-3\ln(2) \, \frac{\mu \eetil^2 - 8\,(\uc-1)}{\mu \eetil^2 - 8\,(\uc-1)\,n \mu}}.
    \end{align*}
    By inequality~\eqref{eq:eeIneq} we get $\mu \eetil^2 - 8\,(\uc-1)\,n \mu > 0$ and since $\mu \ge 1/n$ we have $\mu \eetil^2 - 8\,(\uc-1) \ge \mu \eetil^2 - 8\,(\uc-1)\,n \mu$. Hence,
    \[
        \Pr[M < k_1] \le e^{-3\ln(2)} = \tfrac{1}{8}.
    \]
    We analogously get $\Pr[M > k_2] \le \tfrac{1}{8}$.  Plugging both results in \eq{FPRASproof:suffices}
    finishes the proof.
\end{proof}


\section{Volume computation of intersections}
\label{sec:intersection}

In this section we are considering the complement to the union problem.
We show that surprisingly the volume of a intersection of a set of bodies is often much
harder to calculate than its union.  For many classes of geometric objects there
is even no randomized approximation possible.

As the problem of computing the volume of a polytope is
\SPhard~\cite{DyerF88,Khachiyan89}, so is the
computation of the volume of the intersection of general (i.e., not necessarily
axis-parallel) boxes in~$\Rd$.
This can be seen by describing a polytope as an
intersection of halfplanes and representing these
as general boxes.

Let us now consider the convex bodies again. Trivially, the intersection of
convex bodies is convex itself.  From the oracles defining the given bodies
$B_1,\ldots,B_n$ one can simply construct an oracle which answers
\PointQuery for the intersection of those objects: Given a point $x \in
\Rd$ it asks all $n$ oracles and returns true if and only if $x$ lies in all the bodies.
One could now believe that we can apply the result of \citet{DyerFK91} and the
subsequent improvements mentioned in the introduction to approximate the volume
of the intersection and get an FPRAS for the problem at hand. The problem with
this approach is that the intersection is not ``well-guaranteed.''
That is, there is no point
known that lies in the intersection, not to speak of a sphere inside it.
However, the algorithm of \citet{DyerFK91} relies vitally on the
assumption that the given body is well-guaranteed and hence
cannot be applied for approximating
the volume of the intersection of convex bodies.

We will now present a hard subproblem which shows that the volume of
the intersection cannot be approximated (deterministic or randomized) in
general.
\begin{definition}
    \label{def:B}
    For $p\in[0,1]^d$, let
    $B_{p}:=\{ x \mid 0\leq x_i\leq p_i \ \forall 1 \le i \le d\}$. We call
    $\oB_{p}:=[0,1]^d \setminus B_p$ a \emph{co-box}.
\end{definition}
A co-box is a box where we cut out another box at one corner.
The resulting object can itself be a box, too, but in general it is not even convex.
It can be seen as the complement of a box $B_{p}$ relative to a larger background box $[0,1]^d$.
Note that it is easy to calculate the union of a set of co-boxes $\{\oB_{p_1},\oB_{p_2},\ldots,\oB_{p_n}\}$
with $p_1,p_2,\ldots,p_n\in[0,1]^d$ as
$\bigcup \oB_{p_i} = [0,1]^d \setminus \bigcap B_{p_i}$.
On the other hand, the calculation of the intersection of a set of co-boxes
is \SPhard by \thmref{hardness} as $\bigcup B_{p_i} = [0,1]^d \setminus \bigcap \oB_{p_i}$.
The following theorem shows that it is not even approximable.
\begin{theorem}
    \label{thm:noapprox}
    Let $p_1,p_2,\ldots,p_n \in\Rdp$. Then the volume of \ $\bigcap_{i=1}^n \oB_{p_i}$
    cannot be approximated in (deterministic or randomized) polynomial time
    by a factor of $2^{d^{1-\ee}}$ for any $\ee > 0$ unless $\NP = \BPP$.
\end{theorem}
\begin{proof}
    Consider again the problem \SMONCNF already defined in \secref{union}.
    Let $f = \bigwedge_{k=1}^n \bigvee_{i \in C_k} x_i$
    be a monotone Boolean formula given in CNF as defined in the proof of \thmref{hardness}.
    We now construct for every clause $C_k$ a
    co-box $\oB_{p_k}$ with 
    $p_k = (p_1^{(k)},\ldots,p_d^{(k)})$
    and $p_i^{(k)} = 1/2$ if $i \in C_k$,
    and $p_i^{(k)} = 1$ otherwise.
    The boxes $B_{p_k}$ here correspond to the $A_k$ in the proof of \thmref{hardness}.
    For $x=(x_1,\ldots,x_d)\in\{0,1\}^d$ let
    $U_x=\big[\frac{1}{2} x_1, \frac{1}{2}(x_1+1)\big]\times \cdots
        \times \big[\frac{1}{2}x_d,\frac{1}{2}(x_d+1)\big]$.
    As in the proof of \thmref{hardness} one shows that
    $U_x \subseteq \bigcup_{k=1}^n B_{p_k}$ if and only if $x$ satisfies $\bar{f}$
    and hence also $U_x \subseteq \bigcap_{k=1}^n \oB_{p_k}$ if and only if $x$ satisfies $f$.
    This implies that the volume of $\bigcap_{i=1}^n \oB_{p_i}$ times $2^d$
    equals the number of
    satisfying assignments of $f$.
    \citet{Roth96} showed that \SMONCNF cannot be approximated by
    a factor of $2^{d^{1-\ee}}$ unless $\NP = \BPP$.
    By above reduction, the same inapproximability must hold for
    the volume of $\bigcap_{i=1}^n \oB_{p_i}$.
\end{proof}

This shows that in general there does not exist a polynomial time multiplicative \epsapproximation of the volume
of the intersection of bodies in~$\Rd$.  This holds for all classes of objects which include co-boxes,
e.g. schlicht domains (cf.~\secref{otherobj}).
Though there is no multiplicative approximation, we can still give an
additive approximation algorithm, that is, we can efficiently find a number $\tilde{V}$ such that
\[
    \Pr[V - \ee \, V_{\min} \le \tilde{V} \le V + \ee \, V_{\min}] \ge \tfrac{3}{4}
\]
where $V$ is the exact volume of the intersection and $V_{\min}$ is the minimal
volume of any of the given bodies $B_1,\ldots,B_n$. If we could replace
$V_{\min}$ by $V$ in the equation above, we would have an FPRAS.
This is not possible in general as the ratio of $V$ and $V_{\min}$ can be arbitrarily small.
Hence, such an \epsapproximation is not relative to the exact result, but to the
volume of some greater body.  This is an additive approximation since after
rescaling, so that $V_{\min} \le 1$ we get an additive error of $\ee$. Clearly,
we get the result from above by uniform sampling in the body
$B_{\min}$ corresponding to the volume $V_{\min}$.
Consider $\tilde{V} = V_{\min} \,
( Z_1 + \ldots + Z_N ) / N$, where $Z_i$ is a random variable valued 1, if the
$i$\nobreakdash-th sample point $x_i = \SampleQuery(B_{\min})$ lies in the
intersection of $B_1,\ldots,B_n$, and 0 otherwise.
Using Chebyshev's inequality one can show quite easily that $\tilde{V}$ gives an approximation
as desired, if we choose $N$ proportional to $1/\ee^2$ with the right factor.
This gives an approximation algorithm with runtime
$\Ohbig{n V(d) + \frac{1}{\ee^2} S(d) + \frac{n}{\ee^2} P(d)}$,
yielding $\Ohbig{\frac{d n}{\ee^2}}$ for (not necessarily axis-parallel) boxes.


\section{Discussion and open problems}
\label{sec:disc}

\begin{table*}[t]
    \begin{center}
    \begin{tabularx}{\linewidth}{@{\ \ }l>{\centering}X>{\centering\let\\=\tabularnewline}X}
    \hline
    \hline
    \bf Geometric objects & \bf Volume of the union & \bf Volume of the intersection\\
    \hline
    axis-parallel boxes &
    \SPhard + FPRAS &
    easy \\
    general boxes &
    \SPhard + FPRAS &
    \SPhard \\
    co-boxes &
    easy &
    \SPhard + \APXhard\footnotemark[6]\nphantom{\footnotemark[6]} \\
    schlicht domains &
    \SPhard + FPRAS\footnotemark[5]\nphantom{\footnotemark[5]} &
    \SPhard + \APXhard\footnotemark[6]\nphantom{\footnotemark[6]} \\
    convex bodies &
    \SPhard + FPRAS  &
    \SPhard \\
    \hline
    \hline
    \end{tabularx}
    \end{center}
    \caption{Results for the computational complexity of the calculation of
             the volume of union and intersection (asymptotic in the dimension $d$).
             }
    \label{tab:tab}
\end{table*}

\footnotetext[5]{If the integrals are computable in polynomial time (cf.\ \secref{otherobj}).}
\footnotetext[6]{\thmref{noapprox} even proves that for every fixed $\ee>0$
    approximating the volume within $2^{d^{1-\ee}}$ is \NPhard.}

We have proven \SPhardness for the exact computation of the volume of the union
of bodies in~$\Rd$ as long as the class of bodies includes axis-parallel boxes.
The same holds for the intersection if the class of bodies contains general boxes.
We have also presented an FPRAS for approximating the volume of the union of bodies
that allow three very natural oracles.
Very recently, there appeared a few deterministic polynomial-time approximations
(FPTAS) for hard counting problems
(e.g.\ \citep{HalmanKLOS08,HalmanKLOS08,BandyopadhyayG06,BayatiGKNT07,GamarnikK07,Weitz06}).
It seems to be a very interesting open question whether there exists a deterministic approximation
for the union of some non-trivial class of bodies.
Since the volume of convex bodies determined by oracles
cannot be approximated to within a factor that is
exponential in~$d$~\citep{BaranyF87},
the existence of such a deterministic
approximation for the union seems implausible.
It is also open whether there is a constant~$C$ so that \KMP can be
efficiently deterministically approximated within a factor of~$C$, i.e., if they are in \APX?

For the intersection we proved that no multiplicative approximation (deterministic or
randomized) is possible for co-boxes (cf.\ \defref{B}),
but we also presented a very simple additive approximation algorithm for the intersection problem.
It would be interesting to know if there is a hard class for multiplicative approximation
which contains only convex bodies.

Our results are summarized in \tabref{tab}.
Note the correspondence between axis-parallel boxes and co-boxes.
The discrete counterpart to their approximability and inapproximability
is the approximability of \SDNF
and the inapproximability of \SSAT.

\section*{Acknowledgements}

We thank Joshua Cooper for suggesting the proof of \thmref{hardness},
Ernst Albrecht for discussions on schlicht domains, and
Eckart Zitzler for comments on an earlier version of this paper.
This work was partially supported by a postdoctoral fellowship
from the German Academic Exchange Service (DAAD).



\begin{thebibliography}{30}
\providecommand{\natexlab}[1]{#1}
\providecommand{\url}[1]{\texttt{#1}}
\expandafter\ifx\csname urlstyle\endcsname\relax
  \providecommand{\doi}[1]{doi: #1}\else
  \providecommand{\doi}{doi: \begingroup \urlstyle{rm}\Url}\fi

\bibitem[Agarwal et~al.(2007)Agarwal, Kaplan, and Sharir]{AgarwalKS07}
P.~K. Agarwal, H.~Kaplan, and M.~Sharir.
\newblock Computing the volume of the union of cubes.
\newblock In \emph{Proc.\ 23rd annual Symposium on Computational Geometry
  (SoCG~'07)}, pp. 294--301, 2007.

\bibitem[Applegate and Kannan(1991)]{ApplegateK91}
D.~Applegate and R.~Kannan.
\newblock Sampling and integration of near log-concave functions.
\newblock In \emph{Proc.\ 23rd Annual ACM Symposium on Theory of Computing
  (STOC~'91)}, pp. 156--163, 1991.

\bibitem[Bandyopadhyay and Gamarnik(2006)]{BandyopadhyayG06}
A.~Bandyopadhyay and D.~Gamarnik.
\newblock Counting without sampling: new algorithms for enumeration problems
  using statistical physics.
\newblock In \emph{Proc.\ 17th Annual ACM-SIAM Symposium on Discrete Algorithms
  (SODA~'06)}, pp. 890--899, 2006.

\bibitem[B{\'a}r{\'a}ny and F{\"u}redi(1987)]{BaranyF87}
I.~B{\'a}r{\'a}ny and Z.~F{\"u}redi.
\newblock Computing the volume is difficult.
\newblock \emph{Discrete {\&} Computational Geometry}, 2:\penalty0 319--326,
  1987.

\bibitem[Bayati et~al.(2007)Bayati, Gamarnik, Katz, Nair, and
  Tetali]{BayatiGKNT07}
M.~Bayati, D.~Gamarnik, D.~A. Katz, C.~Nair, and P.~Tetali.
\newblock Simple deterministic approximation algorithms for counting matchings.
\newblock In \emph{Proc.\ 39th Annual ACM Symposium on Theory of Computing
  (STOC~'07)}, pp. 122--127, 2007.

\bibitem[Bentley(1977)]{Bentley77}
J.~L. Bentley.
\newblock Algorithms for {Klee's} rectangle problems, 1977.
\newblock Department of Computer Science, Carnegie Mellon University,
  Unpublished notes.

\bibitem[Bringmann and Friedrich(2008)]{BF08}
K.~Bringmann and T.~Friedrich.
\newblock Approximating the volume of unions and intersections of
  high-dimensional geometric objects.
\newblock In \emph{Proc.\ 19th International Symposium on Algorithms and
  Computation (ISAAC~'08)}, Vol. 5369 of \emph{LNCS}, pp. 436--447, 2008.

\bibitem[Chan(2003)]{Chan03}
T.~M. Chan.
\newblock Semi-online maintenance of geometric optima and measures.
\newblock \emph{SIAM J. Comput.}, 32:\penalty0 700--716, 2003.

\bibitem[Chan(2010)]{Chan09}
T.~M. Chan.
\newblock A (slightly) faster algorithm for klee's measure problem.
\newblock \emph{Computational Geometry}, 43:\penalty0 243 -- 250, 2010.

\bibitem[Chandrasekaran et~al.(2010)Chandrasekaran, Dadush, and Vempala]{CDV10}
K.~Chandrasekaran, D.~Dadush, and S.~Vempala.
\newblock Thin partitions: {Isoperimetric} inequalities and a sampling
  algorithm for star shaped bodies.
\newblock In \emph{Proc.\ 21th Annual ACM-SIAM Symposium on Discrete Algorithms
  (SODA~'10)}, pp. 1630--1645, 2010.

\bibitem[Dyer and Frieze(1988)]{DyerF88}
M.~E. Dyer and A.~M. Frieze.
\newblock On the complexity of computing the volume of a polyhedron.
\newblock \emph{SIAM J.\ Comput.}, 17:\penalty0 967--974, 1988.

\bibitem[Dyer et~al.(1991)Dyer, Frieze, and Kannan]{DyerFK91}
M.~E. Dyer, A.~M. Frieze, and R.~Kannan.
\newblock A random polynomial time algorithm for approximating the volume of
  convex bodies.
\newblock \emph{J. ACM}, 38:\penalty0 1--17, 1991.

\bibitem[Fredman and Weide(1978)]{FredmanW78}
M.~L. Fredman and B.~W. Weide.
\newblock On the complexity of computing the measure of $\bigcup[a_i, b_i]$.
\newblock \emph{Commun. ACM}, 21:\penalty0 540--544, 1978.

\bibitem[Gamarnik and Katz(2007)]{GamarnikK07}
D.~Gamarnik and D.~Katz.
\newblock Correlation decay and deterministic {FPTAS} for counting
  list-colorings of a graph.
\newblock In \emph{Proc.\ 18th Annual ACM-SIAM Symposium on Discrete Algorithms
  (SODA~'07)}, pp. 1245--1254, 2007.

\bibitem[Halman et~al.(2008)Halman, Klabjan, Li, Orlin, and
  Simchi-Levi]{HalmanKLOS08}
N.~Halman, D.~Klabjan, C.-L. Li, J.~B. Orlin, and D.~Simchi-Levi.
\newblock Fully polynomial time approximation schemes for stochastic dynamic
  programs.
\newblock In \emph{Proc.\ 19th Annual ACM-SIAM Symposium on Discrete Algorithms
  (SODA~'08)}, pp. 700--709, 2008.

\bibitem[Kannan et~al.(1997)Kannan, Lov{\'a}sz, and Simonovits]{KannanLS97}
R.~Kannan, L.~Lov{\'a}sz, and M.~Simonovits.
\newblock Random walks and an {$O^*(n^5)$} volume algorithm for convex bodies.
\newblock \emph{Random Struct. Algorithms}, 11:\penalty0 1--50, 1997.

\bibitem[Kaplan et~al.(2007)Kaplan, Rubin, Sharir, and Verbin]{KaplanRSV07}
H.~Kaplan, N.~Rubin, M.~Sharir, and E.~Verbin.
\newblock Counting colors in boxes.
\newblock In \emph{Proc.\ 18th Annual ACM-SIAM Symposium on Discrete Algorithms
  (SODA~'07)}, pp. 785--794, 2007.

\bibitem[Karp et~al.(1989)Karp, Luby, and Madras]{KarpLM89}
R.~M. Karp, M.~Luby, and N.~Madras.
\newblock Monte-carlo approximation algorithms for enumeration problems.
\newblock \emph{J. Algorithms}, 10:\penalty0 429--448, 1989.

\bibitem[Khachiyan(1989)]{Khachiyan89}
L.~G. Khachiyan.
\newblock The problem of calculating the volume of a polyhedron is enumerably
  hard.
\newblock \emph{Russian Mathematical Surveys}, 44:\penalty0 199--200, 1989.

\bibitem[Klee(1977)]{Klee77}
V.~Klee.
\newblock Can the measure of {$\bigcup[a_i, b_i]$} be computed in less than
  {$O(n \log n)$} steps?
\newblock \emph{American Mathematical Monthly}, 84:\penalty0 284--285, 1977.

\bibitem[Lov{\'a}sz and Simonovits(1990)]{LovaszS90}
L.~Lov{\'a}sz and M.~Simonovits.
\newblock The mixing rate of markov chains, an isoperimetric inequality, and
  computing the volume.
\newblock In \emph{Proc.\ 31st Annual Symposium on Foundations of Computer
  Science (FOCS~'90)}, pp. 346--354, 1990.

\bibitem[Lov{\'a}sz and Simonovits(1993)]{LovaszS93}
L.~Lov{\'a}sz and M.~Simonovits.
\newblock Random walks in a convex body and an improved volume algorithm.
\newblock \emph{Random Struct. Algorithms}, 4:\penalty0 359--412, 1993.

\bibitem[Lov{\'a}sz and Vempala(2006)]{LovaszV06}
L.~Lov{\'a}sz and S.~Vempala.
\newblock Simulated annealing in convex bodies and an {$O^*(n^4)$} volume
  algorithm.
\newblock \emph{J. Comput. Syst. Sci.}, 72:\penalty0 392--417, 2006.

\bibitem[Muller(1959)]{Muller59}
M.~E. Muller.
\newblock A note on a method for generating points uniformly on $n$-dimensional
  spheres.
\newblock \emph{Commun. ACM}, 2:\penalty0 19--20, 1959.

\bibitem[Overmars and Yap(1991)]{OvermarsY91}
M.~H. Overmars and C.-K. Yap.
\newblock New upper bounds in {Klee's} measure problem.
\newblock \emph{SIAM J. Comput.}, 20:\penalty0 1034--1045, 1991.

\bibitem[Roth(1996)]{Roth96}
D.~Roth.
\newblock On the hardness of approximate reasoning.
\newblock \emph{Artif. Intell.}, 82:\penalty0 273--302, 1996.

\bibitem[Suzuki and Ibaraki(2004)]{SuzukiI04}
S.~Suzuki and T.~Ibaraki.
\newblock An average running time analysis of a backtracking algorithm to
  calculate the measure of the union of hyperrectangles in $d$ dimensions.
\newblock In \emph{Proc.\ 16th Canadian Conference on Computational Geometry
  (CCCG~'04)}, pp. 196--199, 2004.

\bibitem[van Leeuwen and Wood(1981)]{LeeuwenW81}
J.~van Leeuwen and D.~Wood.
\newblock The measure problem for rectangular ranges in $d$-space.
\newblock \emph{J.\ Algorithms}, 2:\penalty0 282--300, 1981.

\bibitem[Weitz(2006)]{Weitz06}
D.~Weitz.
\newblock Counting independent sets up to the tree threshold.
\newblock In \emph{Proc.\ 38th Annual ACM Symposium on Theory of Computing
  (STOC~'06)}, pp. 140--149, 2006.

\bibitem[Zitzler and Thiele(1999)]{ZitzlerT99}
E.~Zitzler and L.~Thiele.
\newblock Multiobjective evolutionary algorithms: a comparative case study and
  the strength {Pareto} approach.
\newblock \emph{IEEE Trans.\ Evolutionary Computation}, 3:\penalty0 257--271,
  1999.

\end{thebibliography}
\end{document}